\documentclass[aps,prd, preprintnumbers, twocolumn, floats, tightenlines, superscriptaddress]{revtex4}
\usepackage{amsmath,amssymb,amsfonts}
\usepackage{graphicx}
\usepackage{enumerate} 
\usepackage{colordvi} 
\usepackage{color} 
\usepackage{bm}

\newcommand{\bfv}{\mbox{\boldmath$v$}}

\newcommand{\bfk}{\mbox{\boldmath$k$}}
\newcommand{\bfp}{\mbox{\boldmath$p$}}
\newcommand{\bfq}{\mbox{\boldmath$q$}}
\newcommand{\bfr}{\mbox{\boldmath$r$}}
\newcommand{\bfs}{\mbox{\boldmath$s$}}

\newcommand{\bfz}{\mbox{\boldmath$z$}}

\newcommand{\sigmad}{\sigma_{\rm d}}
\newcommand{\sigmav}{\sigma_{\rm v}}
\newcommand{\rhom}{\rho_{\rm m}}

\begin{document}
\title{Constructing perturbation theory kernels for large-scale structure in generalized cosmologies}

\author{Atsushi Taruya}
\affiliation{Center for Gravitational Physics, Yukawa Institute for Theoretical Physics, Kyoto University, Kyoto 606-8502, Japan}
\affiliation{
Kavli Institute for the Physics and Mathematics of the Universe, Todai Institutes for Advanced Study, the University of Tokyo, Kashiwa, Chiba 277-8583, Japan (Kavli IPMU, WPI)}
%
\date{\today}
\begin{abstract}
We present a simple numerical scheme for perturbation theory (PT) calculations of large-scale structure. Solving the evolution equations for perturbations numerically, we construct the PT kernels as building blocks of statistical calculations, from which the power spectrum and/or correlation function can be systematically computed. The scheme is especially applicable to the generalized structure formation including modified gravity, in which the analytic construction of PT kernels is intractable. As an illustration, we show several examples for power spectrum calculations in $f(R)$ gravity and $\Lambda$CDM models. 
\end{abstract}

\pacs{98.80.-k, 98.62.Py, 98.65.-r}
\keywords{cosmology, large-scale structure}
\preprint{YITP-16-69}
\maketitle

\section{Introduction}

Evolution of large-scale matter inhomogeneities is driven by gravity in a cold dark matter dominated universe. As decreasing redshift, development of gravitational clustering eventually enters the nonlinear regime, and the linear theory prediction ceases to be adequate. Even at large scales, the transition to nonlinear evolution appears as a non-negligible effect, which has to be properly incorporated into theoretical predictions in confronting with precision observations. Indeed, aiming at measuring the baryon acoustic oscillations (BAO) and redshift-space distortions (RSD) as probes of the cosmic expansion and growth of structure (e.g., \cite{Jain:2013wgs,Huterer:2013xky,2013PhR...530...87W}), an accuracy of theoretical calculation needs to be better controlled, including observational systematics as well as new cosmological effects such as modification of gravity or free-streaming suppression of massive neutrinos.

Among several approaches to deal with nonlinear structure formation, perturbation theory (PT) of large-scale structure is suited for predicting statistical quantities at cosmological scales of our interest \cite{Bernardeau:2001qr}, particularly relevant for BAO and RSD. Albeit its limitation to weakly nonlinear regime, the PT treatment tells us how the nonlinear clustering is developed through the coupling between different Fourier modes in analytic way, characterized by the so-called (standard) PT kernels. In particular, in the standard cosmological model, the PT kernels are systematically constructed with recursion relations (e.g., \cite{Goroff:1986ep,Bernardeau:2001qr,Crocce:2005xy}), and higher-order corrections to the power spectrum or bispectrum are computed efficiently (e.g., \cite{Makino:1991rp,Jain:1993jh,Scoccimarro:1995if, Scoccimarro:1996se,Scoccimarro:1996jy}). Further, these PT kernels are applied to several resummed PT calculations recently developed (e.g., \cite{Crocce:2005xy,Crocce:2005xz,Crocce:2007dt,Matsubara:2007wj,Taruya:2007xy,Crocce:2012fa,Bernardeau:2008fa,Bernardeau:2011dp,Bernardeau:2012ux,Taruya:2012ut}), with which the applicable range of PT prediction becomes greatly improved.

One important remark in the present PT treatment is, however, that the calculations heavily rely on the analytic PT kernels constructed with recursion relations. Apart from a few exceptional case including the Einstein-de Sitter universe, the analytic construction of PT kernels is generally intractable. This is even the cases for standard Lambda cold dark matter ($\Lambda$CDM) model, in which the mode coupling successively generates a set of higher-order growth functions, and a tractable higher-order calculation needs to be handled by the so-called Einstein-de Sitter approximation (e.g., Ref.~\cite{Takahashi:2008yk,Pietroni:2008jx,2009PhRvD..79j3526H}). Although some of the cases are cured by generalizing the analytic recursion relation \cite{Fasiello:2016qpn}, more difficult cases arise from the modification of gravity or structure formation. An example is the modified gravity models with the Chameleon-type screening mechanisms \cite{Khoury:2003rn} (e.g., $f(R)$ gravity \cite{Starobinsky:2007hu,Hu:2007nk}). In these models, gravity sector is modified in the presence of a new scalar degree of freedom, coupled with Poisson equation. The resultant evolution of perturbations are not separately treated in time and scales, and the time evolution of statistical quantities has to be numerically solved (e.g., \cite{Koyama:2009me}).

In order to deal with PT calculation in an analytically intractable situation, 
several numerical approaches have been so far presented (e.g., \cite{Valageas:2006bi,Pietroni:2008jx,Koyama:2009me,2009PhRvD..79j3526H,Carlson:2009it}). These approaches are basically the moment-based method that numerically solves the time evolution of statistical quantities. The methods particularly rely on a specific resummed PT formalism, and give a statistical prediction in a wider applicable range.

In this paper, rather than computing the statistical quantities directly, we are interested in the numerical scheme to reconstruct the PT kernels, since these are building blocks of various perturbative treatments, and many applications would be possible. We here present a simple numerical method to reconstruct the PT kernels. As an illustration, we consider the structure formation in $f(R)$ gravity model as well as general relativity (GR), and show several examples for power spectrum calculations based on both the standard PT and resummed PT treatments.

This paper is organized as follows. In Sec.~\ref{sec:BasicEqs_MG}, we begin by briefly reviewing the basic equations for perturbations based on a general framework to deal with a wide class of modified gravity models. Sec.~\ref{sec:PT_kernel} then describes the numerical treatment to reconstruct the PT kernels. With a relevant cosmological setup and initial conditions, PT kernels are numerically reconstructed by solving the evolution equations for perturbations. Sec.~\ref{sec:demonstration} demonstrates the numerical scheme for PT kernels, and presents several examples for the power spectrum calculation, based on the standard and resummed PT treatment in both GR and $f(R)$ gravity. Finally, Sec.~\ref{sec:summary} is devoted to summary and discussion. 

\section{Basic equations for perturbations}
\label{sec:BasicEqs_MG}

In this section, we present a framework to deal with PT calculations of large-scale structure. We are particularly interested in a wide class of structure formation that can differ from the standard $\Lambda$CDM model. In this paper, we consider the PT formalism developed by Ref.~\cite{Koyama:2009me}, with which we can describe the structure formation in a variety of modified gravity models that has a nonlinear screening mechanism to recover GR.

On large scales of our interest,  where the dynamics of matter fluctuations is approximately described by the single-stream approximation of collisionless Boltzmann equation, the evolution of CDM plus baryon system can be 
regarded as an irrotational and pressureless fluid system. The governing equations for matter fluctuations become
\begin{align}
&\frac{\partial \delta}{\partial t} +
\frac{1}{a}\nabla\cdot[(1+\delta)\bfv]=0,
\label{eq:eq_continuity}\\
&\frac{\partial \bfv}{\partial t} + H\,\bfv+
\frac{1}{a}(\bfv\cdot\nabla)\cdot\bfv=-\frac{1}{a}\nabla\psi,
\label{eq:eq_Euler}
\end{align}
where $\psi$ is the Newton potential. In modified gravity, the gravity sector relevant for structure formation is generally modified in the presence of new scalar degree of freedom. In most of the models, this modification can be described by the Brans-Dicke type scalar field, and the Newton potential couples with both the matter and Brans-Dicke scalar. The modified Poisson equation must be solved with scalar-field equation:  
\begin{align}
&\frac{1}{a}\nabla^2\psi=\frac{\kappa^2}{2}\,\rhom\,\delta
-\frac{1}{2a^2}\nabla^2\varphi,
\label{eq:Poisson_eq}\\
&(3+2\omega_{\rm BD})\frac{1}{a^2}\nabla^2\varphi=-2\kappa^2\rhom\,\delta
-\mathcal{I}(\varphi)
\label{eq:EoM_scalaron}
\end{align}
with $\kappa^2=8\pi\,G$ and $\omega_{\rm BD}$ being the Brans-Dicke parameter.
Here, we employ the quasi-static approximation, valid at the sub-horizon scales. Note that the field $\varphi$ has a nonlinear self-interaction term, 
$\mathcal{I}$, by which the screening mechanisms that recovers GR at nonlinear  
regime can be realized. In PT framework, it is expanded as 
\begin{align}
&\mathcal{I}(\varphi)=M_1(k)
\nonumber\\
&+\frac{1}{2}\,
\int\frac{d^3\bfk_1 d^3\bfk_2}{(2\pi)^3}\delta_{\rm D}(\bfk-\bfk_{12})\,
M_2(\bfk_1,\bfk_2)\varphi(\bfk_1)\varphi(\bfk_2)
\nonumber\\
&+\frac{1}{6}\,
\int\frac{d^3\bfk_1 d^3\bfk_2d^3\bfk_3}{(2\pi)^6}\delta_{\rm D}(\bfk-\bfk_{123})\,
M_3(\bfk_1,\bfk_2,\bfk_3)
\nonumber\\
&\qquad\quad\times\varphi(\bfk_1)\varphi(\bfk_2)\varphi(\bfk_3)+\cdots
\label{eq:I_expansion}
\end{align}
The functions $M_n$ are in general model-dependent, and are explicitly 
given provided the concrete model of modified gravity \cite{Koyama:2009me}. In this paper, we will demonstrate our numerical PT treatment described in Sec.~\ref{sec:PT_kernel}, specifically focusing on GR and $f(R)$ gravity. The functions $M_n$ are then explicitly given by \cite{Koyama:2009me}
\begin{align}
&M_n=\left\{\begin{array}{cl}
0, & \,\,\mbox{GR}
\\
\\
{\displaystyle \frac{d^n\overline{R}(f_R)}{d\, f_R^n}}, & \,\,f(R)\,\,\mbox{gravity}
\end{array}\right.
\label{eq:func_Mn}
\end{align}
with $\overline{R}$ being the background curvature and $f_R\equiv df(R)/dR$. 
As a successful $f(R)$ gravity model for the late-time cosmology, we consider the specific functional form in the high-curvature limit, given by $f(R)= -2\kappa^2\,\rho_\Lambda+|f_{\rm R,0}|\,(R_0^2/R)$ with $R_0$ being background curvature today. This model has been frequently studied in the literature, and with $|f_{\rm R,0}|\ll1$, the background cosmic expansion becomes nearly identical to the $\Lambda$CDM model. In this case, Eq.~(\ref{eq:func_Mn}) reduces to $M_n=\{(2n+1)!/(-2\,|f_{\rm R,0}|)^n\} (\overline{R}/R_0)^{2n}\overline{R}$, and we have
$\overline{R}=\kappa^2(\rho_{\rm m}+4\rho_\Lambda)$.  Below, we set the model parameter $|f_{\rm R,0}|$ to $10^{-4}$ for illustrative purpose, and presents the results.

Eqs.~(\ref{eq:eq_continuity})--(\ref{eq:EoM_scalaron}) are the 
basic equations for perturbations. In Fourier space, 
these can be reduced to a more compact form.
Assuming the irrotationality of fluid quantities,
the velocity field is expressed in terms of scalar quantity, 
$\theta=\nabla\cdot\bfv/(aH)$. Then, we have \cite{Koyama:2009me},
\begin{align}
&H^{-1}\frac{\partial \delta(\bfk)}{\partial t}+\theta(\bfk)
\nonumber\\
&\qquad=-\int\frac{d^3\bfk_1d^3\bfk_2}{(2\pi)^3}\,\delta_D(\bfk-\bfk_{12})\,
\alpha(\bfk_1,\bfk_2)\,\theta(\bfk_1)\delta(\bfk_2),
\label{eq:EoM1}
\\
&H^{-1}\frac{\partial \theta(\bfk)}{\partial t}+
\left\{2+\frac{\dot{H}}{H^2}\right\}\theta(\bfk)
+
\frac{\kappa^2\rho_{\rm m}}{2H^2}
\left\{1+\frac{1}{3}\frac{(k/a)^2}{\Pi(\bfk)}\right\}\delta(\bfk)
\nonumber\\
&\qquad
=-\frac{1}{2}\int\frac{d^3\bfk_1d^3\bfk_2}{(2\pi)^3}\,\delta_D(\bfk-\bfk_{12})\,
\beta(\bfk_1,\bfk_2)\,\theta(\bfk_1)\theta(\bfk_2)
\nonumber\\
&\qquad-\frac{1}{2H^2}\left(\frac{k}{a}\right)^2\,S(\bfk).
\label{eq:EoM2}
\end{align}
Here $\alpha$ and $\beta$ are the mode-coupling kernels given by
\begin{align}
&\alpha(\bfk_1,\bfk_2)=1+\frac{\bfk_1\cdot\bfk_2}{|\bfk_1|^2},
\nonumber\\
&\beta(\bfk_1,\bfk_2)=\frac{(\bfk_1\cdot\bfk_2)|\bfk_1+\bfk_2|^2}
{|\bfk_1|^2|\bfk_2|^2}.
\nonumber
\end{align}
The function $\Pi$ characterizes the
deviation of the Newton constant from GR,
while the quantity $S$ is originated from
the non-linear interactions of the scalar field, which is
responsible for the recovery of GR at small scales.
The explicit form of these
are obtained from the Poisson equation and field equation for Brans-Dicke 
scalar [Eqs.~(\ref{eq:Poisson_eq})-(\ref{eq:I_expansion})], 
and the expressions relevant for perturbations up to the third oder are
respectively given by \cite{Koyama:2009me,Taruya:2014faa}:
\begin{align}
&\Pi(k)=\frac{1}{3}\left\{(3+2\omega_{\rm BD})\frac{k^2}{a^2}+M_1(k)\right\},
\end{align}
\begin{widetext}
\begin{align}
&S(k)=-\frac{1}{6\,\Pi(k)} \left(\frac{\kappa^2\,\rho_{\rm m}}{3}\right)^2
\int\frac{d^3\bfk_1d^3\bfk_2}{(2\pi)^3}\,\delta_D(\bfk-\bfk_{12})\,
M_2(\bfk_1,\bfk_2)\frac{\delta(\bfk_1)\delta(\bfk_2)}{\Pi(k_1)\Pi(k_2)}
\nonumber\\
&\qquad -\frac{1}{18\,\Pi(k)} \left(\frac{\kappa^2\,\rho_{\rm m}}{3}\right)^3
\int\frac{d^3\bfk_1d^3\bfk_2d^3\bfk_3}{(2\pi)^6}\,\delta_D(\bfk-\bfk_{123})\,
\left\{M_3(\bfk_1,\bfk_2,\bfk_3)
-\frac{M_2(\bfk_{12},\bfk_3)M_2(\bfk_1,\bfk_2)}{\Pi(k_{12})}\right\}
\frac{\delta(\bfk_1)\delta(\bfk_2)\delta(\bfk_3)}{\Pi(k_1)\Pi(k_2)\Pi(k_3)}.
\label{eq:func_S}
\end{align}
\end{widetext}
Here, in deriving the last expression,
we perturbatively express the scalar field $\varphi$ in terms of $\delta$
using Eqs.~(\ref{eq:EoM_scalaron}) and (\ref{eq:I_expansion}) (see 
Appendix B of Ref.~\cite{Taruya:2014faa} for derivation).

\section{Solving perturbation theory kernels numerically}
\label{sec:PT_kernel}

The main goal of the PT calculation is to solve Eqs.~(\ref{eq:EoM1}) and (\ref{eq:EoM2}) perturbatively, and to apply their perturbative solutions to the statistical predictions of large-scale structure. To start with, let us expand the quantities $\delta$ and $\theta$ as
\begin{align}
&\delta(\bfk;t) = \delta^{(1)}(\bfk;t) + \delta^{(2)}(\bfk;t) + \cdots, 
\\
&\theta(\bfk;t) = \theta^{(1)}(\bfk;t) + \theta^{(2)}(\bfk;t) + \cdots. 
\end{align}
Our focus here is the evolution of matter fluctuations seeded by a tiny density fluctuation. In this case, the solutions of perturbations are expressed as\footnote{This might be regarded as a specific initial condition in the sense that the randomness of the velocity field is determined solely by the initial density field, however, it is relevant for most of the scenarios.}
\begin{align}
&\delta^{(n)}(\bfk;t)= \int\frac{d^3\bfk_1\cdots d^3\bfk_n}{(2\pi)^{3(n-1)}}\,
\delta_{\rm D}(\bfk-\bfk_{12\cdots n})
\nonumber\\
&\qquad\qquad
\times F_n(\bfk_1,\cdots,\bfk_n;\,t)\,\delta_0(\bfk_1)\cdots \delta_0(\bfk_n),
\label{eq:delta_n}\\
&\theta^{(n)}(\bfk;t)= \int\frac{d^3\bfk_1\cdots d^3\bfk_n}{(2\pi)^{3(n-1)}}\,
\delta_{\rm D}(\bfk-\bfk_{12\cdots n})
\nonumber\\
&\qquad\qquad
\times 
G_n(\bfk_1,\cdots,\bfk_n;\,t)\,\delta_0(\bfk_1)\cdots \delta_0(\bfk_n),
\label{eq:theta_n}
\end{align}
where $\delta_0$ is the random initial density fluctuation. The functions $F_n$ and $G_n$ are the so-called standard PT kernels, and in some limited cases, these are analytically constructed with recursions relations \cite{Goroff:1986ep,Bernardeau:2001qr,Crocce:2005xy} based on the Einstein-de Sitter approximation, by which all the non-trivial higher-order growth factors are expressed in terms of the linear growth factor. In general structure formation with Eqs.~(\ref{eq:EoM1}), (\ref{eq:EoM2}), and (\ref{eq:func_S}), however, the systematic construction of PT kernels is analytically intractable (see \cite{Koyama:2009me,Takushima:2015iha} for some attempts in a class of modified gravity models). A typical case is the $f(R)$ gravity, in which the scale- and time-dependence of the perturbation equations are no longer separable.

In this paper, solving the evolution equations, we consider the numerical construction of PT kernels. Defining the linear operator of the matrix form (here $a$ is the scale factor of the Universe): 
\begin{widetext}
\begin{align}
\widehat{\mbox{\boldmath${\mathcal L}$}}(k;\,a)\equiv
\left(
\begin{array}{lll}
{\displaystyle a\frac{d}{da}} &~~ &1 
\\
\\
{\displaystyle 
\frac{3}{2}\left(\frac{H_0}{H(a)}\right)^2 }
{\displaystyle 
\frac{\Omega_{\rm m,0}}{a^3}\,\left\{1+\frac{1}{3}\frac{(k/a)^2}{\Pi(k)}\right\}
}
&~~ &
{\displaystyle a\frac{d}{da}+\left(2+\frac{\dot{H}}{H^2}\right) }
\end{array}
\right),
\end{align}
\label{eq:linear_operator}
the evolution equations for the kernels $F_n$ and $G_n$ are written as
\begin{align}
\widehat{\mbox{\boldmath${\mathcal L}$}}(k_{1\cdots n};\,a)
\left(
\begin{array}{c}
F_n(\bfk_1,\cdots,\bfk_n; a)
\\
\\
G_n(\bfk_1,\cdots,\bfk_n; a)
\end{array}
\right)=\left(
\begin{array}{c}
S_n(\bfk_1,\cdots,\bfk_n; a)
\\
\\
T_n(\bfk_1,\cdots,\bfk_n; a)
\end{array}
\right).
\label{eq:evolution_PTkernel}
\end{align}
\end{widetext}
The source functions $S_n$ and $T_n$ represent the nonlinear mode 
coupling, and are written in terms of the lower-oder perturbed quantities. 
The explicit form of these functions is derived from the basic equations (\ref{eq:EoM1}), (\ref{eq:EoM2}), and (\ref{eq:func_S}), and we below summarize those up to the third order.

\subsection{Source functions}

Obviously, the source function at first-order should vanish, since we do not consider any interaction at linear order. We thus have
\begin{align}
&S_1(k;a)=0,
\label{eq:S_1}\\
&T_1(k;a)=0.
\label{eq:T_1}
\end{align}
At second order, even with the vanishing source terms, the linear-order solution obtained from Eq.~(\ref{eq:evolution_PTkernel}) naturally induces the non-vanishing source function. From Eqs.~(\ref{eq:EoM1}) and (\ref{eq:EoM2}), we can read off
\begin{align}
&S_2(\bfk_1,\bfk_2;a)=
-\frac{1}{2}
\Bigl\{\alpha(\bfk_1,\bfk_2)\,G_1(k_1)F_1(k_2)
\nonumber\\
&\qquad\qquad
+\alpha(\bfk_2,\bfk_1)G_1(k_2)F_1(k_1)\Bigr\},
\label{eq:S_2}\\
&T_2(\bfk_1,\bfk_2;a)=
-\frac{1}{2}\,\beta(\bfk_1,\bfk_2)\,G_1(k_1)G_1(k_2)  
v\nonumber\\
&\qquad\qquad+ 
\frac{1}{12}\left(\frac{k_{12}}{a\,H(a)}\right)^2\frac{H_0^4}{\Pi(k_{12})}
\nonumber\\
&\qquad\qquad\times
\left(\frac{\Omega_{\rm m,0}}{a^3}\right)^2\,M_2(\bfk_1,\bfk_2)\,\frac{F_1(k_1)F_1(k_2)}{\Pi(k_1)\Pi(k_2)}.
\label{eq:T_2}
\end{align}
The source functions given above are expressed in a symmetric form, i.e.,  $S_2(\bfk_1,\bfk_2)=S_2(\bfk_2,\bfk_1)$ and 
$T_2(\bfk_1,\bfk_2)=T_2(\bfk_2,\bfk_1)$. This implies that numerically solving 
Eq.~(\ref{eq:evolution_PTkernel}) with the above source functions automatically gives the symmetrized PT kernel for $F_2$ and $G_2$.

In a similar way,  the third-order source functions are read off from the evolution equations to give
\begin{widetext}
\begin{align}
&S_3(\bfk_1,\bfk_2,\bfk_3;a)= -
\alpha(\bfk_1,\bfk_{23})\,G_1(k_1)F_2(\bfk_2,\bfk_3)
-\alpha(\bfk_{23},\bfk_1)G_2(\bfk_2,\bfk_3)F_1(k_3), 
\label{eq:S_3}
\\
&T_3(\bfk_1,\bfk_2,\bfk_3;a)=-
\beta(\bfk_1,\bfk_{23})\,G_1(k_1)G_2(\bfk_2,\bfk_3)
+\frac{1}{6}\left(\frac{k_{123}}{a\,H(a)}\right)^2\frac{H_0^4}{\Pi(k_{123})}
\left(\frac{\Omega_{\rm m,0}}{a^3}\right)^2\,
M_2(\bfk_1,\bfk_{23})\,\frac{F_1(k_1)F_2(\bfk_2,\bfk_3)}{\Pi(k_1)\Pi(k_{23})}
\nonumber\\
& \qquad\qquad
+\frac{1}{36}\left(\frac{k_{123}}{a\,H(a)}\right)^2\frac{H_0^6}{\Pi(k_{123})}
\left(\frac{\Omega_{\rm m,0}}{a^3}\right)^3\,
\left\{ M_3(\bfk_1,\bfk_2,\bfk_3)-\frac{M_2(\bfk_{23},\bfk_1)M_2(\bfk_2,\bfk_3)}{\Pi(k_{12})}\right\}\frac{F_1(k_1)F_1(k_2)F_1(k_3)}{\Pi(k_1)\Pi(k_2)\Pi(k_3)}.
\label{eq:T_3}
\end{align}
\end{widetext}
Note here that the expressions given above 
are not fully symmetrized with respect to the exchange of each argument, 
but are partly symmetric under $\bfk_2\leftrightarrow\bfk_3$. Thus,  
the PT kernels numerically constructed with the above source functions needs to be properly symmetrized for later analysis in the statistical calculations.  Making use of the partial symmetry, the symmetrized kernels are obtained from
\begin{align}
&F_{3, {\rm sym}}(\bfk_1,\bfk_2,\bfk_3;a)=\frac{1}{3}
\Bigl\{F_3(\bfk_1,\bfk_2,\bfk_3;a)+\mbox{cyclic perm.}\Bigr\}, 
\label{eq:symmetrized_F3}\\
&G_{3, {\rm sym}}(\bfk_1,\bfk_2,\bfk_3;a)=\frac{1}{3}
\Bigl\{G_3(\bfk_1,\bfk_2,\bfk_3;a)+\mbox{cyclic perm.}\Bigr\}.
\label{eq:symmetrized_G3}
\end{align}

\subsection{Initial conditions}
\label{subsec:IC}

We are interested in the structure formation starting with initial condition consitent with CMB observations. In such a case, the Universe 
at an early epoch would be approximately described by the Einstein-de Sitter 
Universe. The evolution of matter fluctuations is dealt with linear theory, from which we obtain the growing-mode solution, $F_1\propto a$ and  $G_1\propto -a$. Since we are also interested in the late-time evolution dominated by the growing mode, as a natural initial condition, we may set 
\begin{align}
F_1(k;a_{\rm i})=a_{\rm i}, \qquad
G_1(k;a_{\rm i})=-a_{\rm i}, 
\label{eq:initial_lin}
\end{align}
where $a_{\rm i}$ is the initial scale factor, which we will typically 
take $a_{\rm i}=10^{-4}$. 
For the higher-order PT kernels, the initial condition becomes
\begin{align}
F_n(\bfk_1,\cdots,\bfk_n;a_{\rm i})=0, \qquad
G_n(\bfk_1,\cdots,\bfk_n;a_{\rm i})=0. 
\label{eq:initial_nth}
\end{align}

While the initial conditions given above may be the most relevant set up consistent with observations, we can of course examine the other setup to test the different structure formation scenarios. As an example, we will present the cases with Zel'dovich initial condition. Note that for statistical calculations, we need to further fix the properties of the initial density field $\delta_0$ in Eqs.~(\ref{eq:delta_n}) and (\ref{eq:theta_n}). In Sec.~\ref{sec:demonstration}, we will demonstrate several examples assuming the Gaussianity of $\delta_0$.

\subsection{Numerical implementation}

Given a set of evolution equations and initial conditions for PT kernels, it is straightforward to obtain numerical solutions. Since the evolution equations for each PT kernel are the ordinary differential equations, and in most of the cases these are expressed in a regular and non-singular form, the standard integrator is sufficient for a precision calculation. We will present below the numerical results based on the Bulirsch-Stoer method (e.g., \cite{Press:1992zz}). 

As it will be demonstrated below, we are particularly interested in the power spectrum calculations at next-to-leading order, called one-loop. In this case, we need at least kernels up to the third order, for which a specific procedure of the numerical calculation is given as follows. Introducing a shortcut notation $\mathcal{F}_n=(F_n,\,G_n)$, for a given set of wave vectors $(\bfk_1,\,\bfk_2,\,\bfk_3)$, 

\begin{enumerate}
\renewcommand{\labelenumi}{(\alph{enumi}).}
\item Solve simultaneously the evolution equations for the kernels,  
$\mathcal{F}_1(k_i)$ $(i=1,2,3)$, 
$\mathcal{F}_2(\bfk_2,\bfk_3)$, and $\mathcal{F}_3(\bfk_1, \bfk_2,\bfk_3)$.

\item Repeat (a) to obtain $\mathcal{F}_3(\bfk_2,\bfk_3,\bfk_1)$ and $\mathcal{F}_3(\bfk_3, \bfk_1,\bfk_2)$. 

\item Combining the results (a) and (b), evaluate the symmetrized kernel, $\mathcal{F}_{3,{\rm sym}}$ through Eq.~(\ref{eq:symmetrized_F3}). 
\end{enumerate}

Note that the kernels $\mathcal{F}_2$ are automatically symmetrized with the source terms in Eqs.~(\ref{eq:S_2}) and (\ref{eq:T_2}). The above procedure is applied to many set of wave vectors used for the mode-coupling (loop) integrals until a sufficient number of kernels are sampled over a wide Fourier modes. 
As we will see in \ref{subsec:SPT}, thank to the statistical isotropy, the data size of the PT kernels needed for power spectrum calculations is not actually so large at one-loop order. The second- and third-order kernels are just tabulated as the three-dimensional array, and hence the kernel data for one-loop calculations can be quickly created even without parallel computation.

\section{Demonstration in power spectrum calculations} 
\label{sec:demonstration}

In this section, numerical scheme to solve PT kernels presented in Sec.~\ref{sec:PT_kernel} is demonstrated in the power spectrum calculations, focusing on both GR and $f(R)$ gravity. We first present the results of standard PT calculation in Sec.~\ref{subsec:SPT}. Application of our numerical treatment to the resummed PT calculation is presented in Sec.~\ref{subsec:resummedPT}. As another interesting application, in Sec.~\ref{subsec:ZAIC}, we examine the power spectrum calculations starting with the Zel'dovich initial condition, and quantify the impact of transients based on the standard PT treatment. Finally, in Sec.~\ref{subsec:RSD}, practical application of our numerical scheme to the modeling of power spectrum in redshift-space is briefly discussed. 

In what follows, we assume Gaussian initial condition, for which the randomness of the initial density field $\delta_0$ is solely characterized by the initial (linear) power spectrum $P_0$: 
\begin{align}
\langle\delta_0(\bfk)\delta_0(\bfk')\rangle=(2\pi)^3\delta_{\rm D}(\bfk+\bfk')\,P_0(k).
\end{align}
Adopting the flat $\Lambda$CDM model, we use the CMB Boltzmann code, $\verb|camb|$ \cite{Lewis:1999bs}, to compute the initial (linear) power spectrum based on the cosmological parameters consistent with nine-year WMAP results \cite{Hinshaw:2012aka}: $\Omega_{\rm m}=0.281$, $\Omega_{\Lambda}=0.719$, $\Omega_{\rm b}=0.0464$, $h=0.697$, $n_{\rm s}=0.971$, $\sigma_8=0.851$ (see Ref.~\cite{Taruya:2014faa}).

\subsection{Standard PT}
\label{subsec:SPT}

\begin{figure}[t]
\hspace*{-1.0cm}
\includegraphics[width=10.7cm]{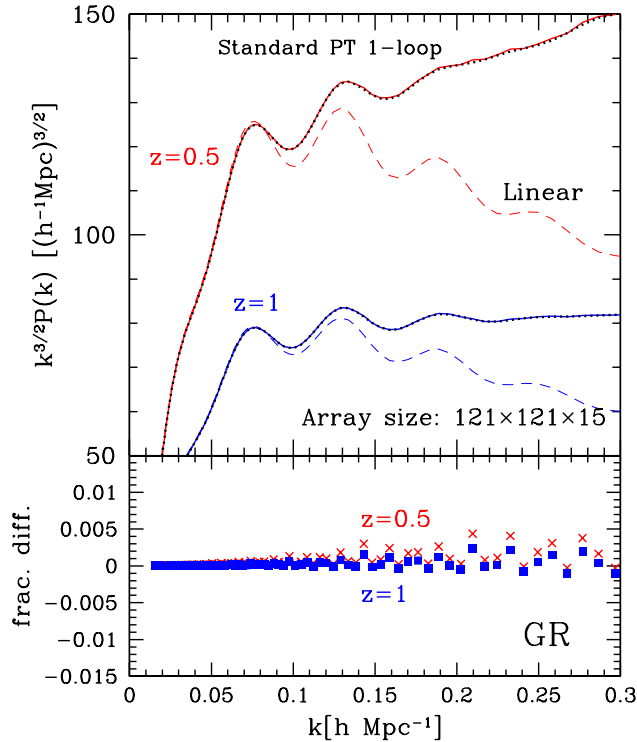}

\vspace*{-0.5cm}

\caption{Matter power spectra at $z=0.5$ (blue) and $z=1$ (red) in standard PT calculations. {\it Top}: power spectrum multiplied by $k^{3/2}$. While solid lines are the one-loop power spectra reconstructed from the numerical data of PT kernels, black dotted lines are the analytic PT results based on the Einstein-de Sitter approximation. For comparison, linear theory predictions are depicted as dashed lines. {\it Bottom}: fractional difference of the standard PT results between numerical approach and analytic treatment, $P^{\rm SPT}_{\rm num}(k)/P^{\rm SPT}_{\rm analytic}(k)-1$. 
\label{fig:pkdiff}}
\end{figure}

Provided the standard PT kernels up to the third order, 
first leading-order corrections called one-loop are computed. 
Here and in what follows, we abbreviate all the symmetrized kernels $F_{n,{\rm sym}}$ to $F_n$. The power spectrum of density field or matter power spectrum, $P_{\delta\delta}$, is given by 
\begin{align}
&P^{\rm SPT}_{\delta\delta}(k)=\{F_1(k)\}^2\,P_0(k) 
\nonumber\\
&\quad + 6\,F_1(k)\int \frac{d^3\bfq}{(2\pi)^3}\,F_3(\bfq,-\bfq,\bfk)\,P_0(q)
\nonumber\\
&\quad +2\,\int\frac{d^3\bfq}{(2\pi)^3}\,\left\{F_2(\bfq,\bfk-\bfq)\right\}^2\,P_0(q)P_0(|\bfk-\bfq|).
\label{eq:pk_dd_spt}
\end{align}
Although the expression apparently involves the three-dimensional integrals, it is known in the GR case that these are reduced to one- and two-dimensional integrals (e.g., \cite{Makino:1991rp,Jain:1993jh,Scoccimarro:1996se}). Recently, a novel algorithm for fast computation has been proposed \cite{Schmittfull:2016jsw,McEwen:2016fjn}. In general structure formation scenarios including modified gravity, such a fast algorithm is no longer adequate, however, statistical isotropy still enables us to reduce these integrals to two-dimensional, which can be quickly evaluated with standard Gaussian quadrature.

Fig.~\ref{fig:pkdiff} shows the power spectra at $z=0.5$ (red) and $1$ (blue) in GR. The results reconstructed from the numerical PT kernels (solid) are compared with those obtained from the analytic kernels (dotted). In computing the power spectrum from the numerical PT kernels,  the kernel data of $F_2(\bfq,\bfk-\bfq)$, $F_3(\bfq,-\bfq,\bfk)$ are first stored in the three-dimensional array $(k,q,\mu)$ with $121$ bins for wavenumbers $k$ and $q$, and with $15$ bins for directional cosine $\mu=(\bfk\cdot\bfq)/(|\bfk||\bfq|)$. The wavenumber $k$ and $q$ are sampled in the range, $[10^{-3},10]\,h$\,Mpc$^{-1}$ in logarithmic scales. With the $121^2\times15$ arrays, it typically costs $50$-$90$ seconds on a laptop computer without parallel computation\footnote{To be precise, timing results are obtained on MacBook Pro with a $2.9$GHz Intel Core i5 processor, using the Intel compiler.}. Then, the stored kernel data $F_2$ and $F_3$ are delivered to the code to compute Eq.~(\ref{eq:pk_dd_spt}), which creates the power spectrum data with typically $10$-$20$ seconds.

The resultant power spectra obtained from the two methods concides with each other, and are indistinguishable. To see the quantitative difference in detail, bottom panel of Fig.~\ref{fig:pkdiff} plots the fractional difference between the two, $[P^{\rm SPT}_{\rm num}(k)-P^{\rm SPT}_{\rm analytic}(k)]/P^{\rm SPT}_{\rm analytic}(k)$, with $P^{\rm SPT}_{\rm num}$ and $P^{\rm SPT}_{\rm analytic}$ being respectively the power spectra computed with numerical and analytic PT kernels. Within the validity range of the standard PT one-loop, which is roughly $k\lesssim 0.1$ and $0.15$ at $z=0.5$ and $1$, the differences are well within $0.1$\%. Extrapolating the results to higher-$k$, there appears a slight systematic increase of the fractional difference. This presumably comes from a small flaw in the power spectrum calculations with analytic PT kernels, for which the Einstein-de Sitter approximation (e.g., Ref.~\cite{Takahashi:2008yk,Pietroni:2008jx,2009PhRvD..79j3526H}) is used to evaluate the higher-order growth factors. Since the systematic deviation arising from this approximation manifests far away from the applicable range of PT, it does not give any impact on the PT calculation at all. Rather, the present numerical scheme is proven to be helpful for a quick check of the analytic PT treatment.

\subsection{Resummed PT calculation}
\label{subsec:resummedPT}

\begin{figure*}[t]
\includegraphics[width=8.9cm]{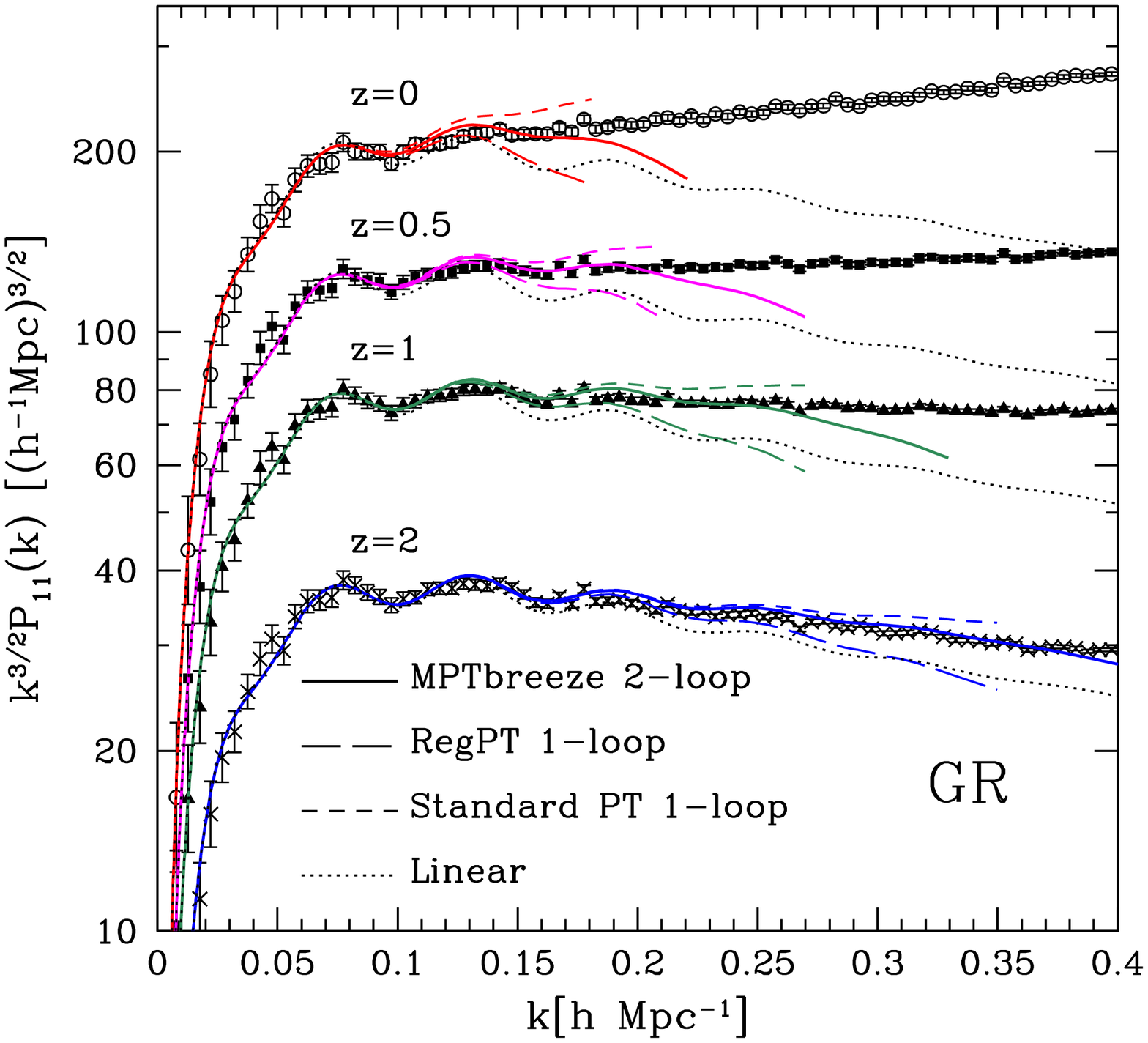}
\includegraphics[width=8.9cm]{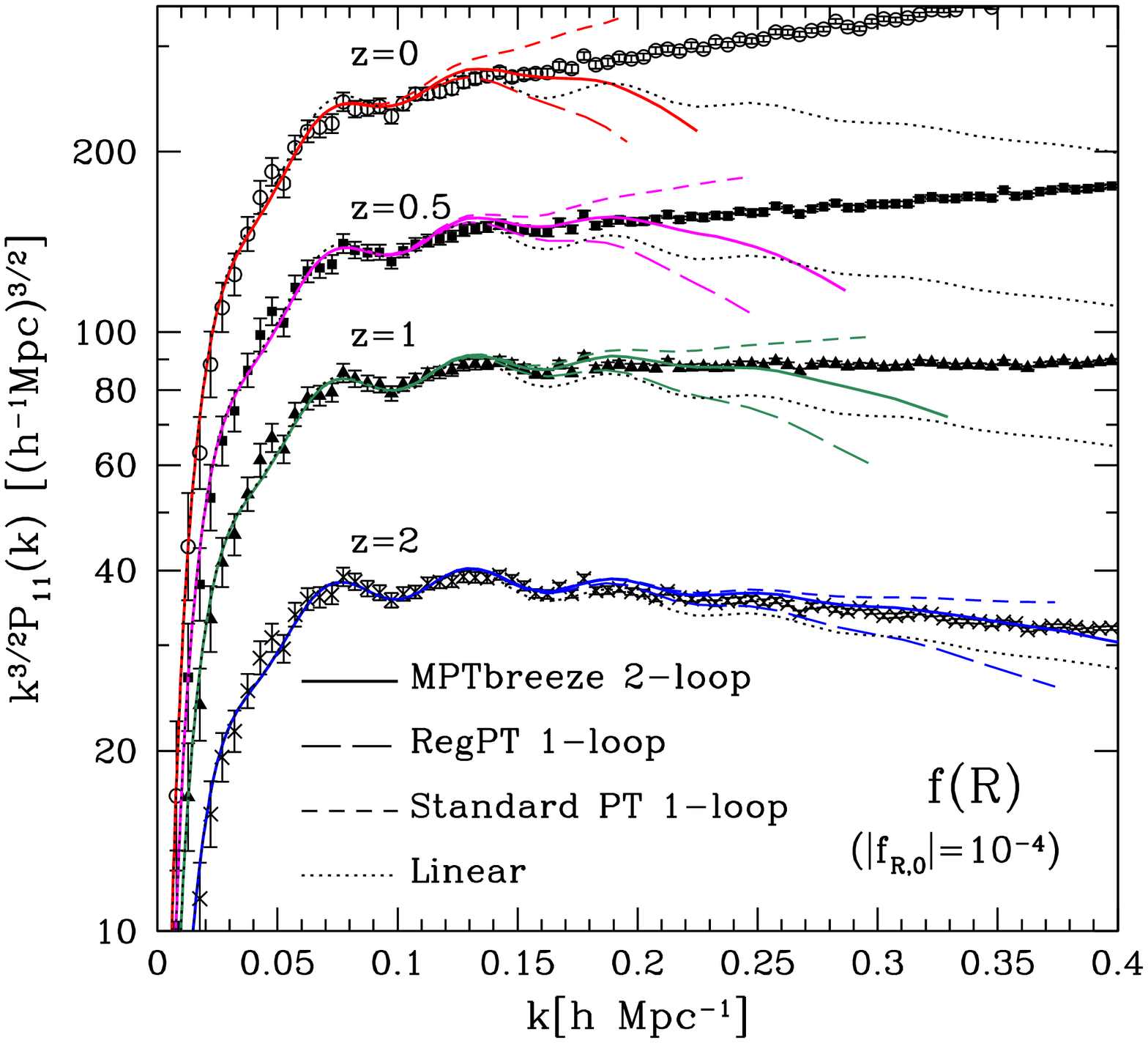}

\vspace*{-0.5cm}

\caption{Matter power spectra in GR (left) and $f(R)$ gravity (right) at $z=0$, $0.5$, $1$, and $2$ (from top to bottom). Based on the numerically constructed PT kernels up to third order, standard and resummed PT calculations are performed, and the results multiplied by $k^{3/2}$ are compared with $N$-body simulations  taken from Ref.~\cite{Taruya:2014faa}. 
Long-dashed short-dashed lines are the one-loop predictions based on RegPT and standard PT, respectively. On the other hand, solid lines represents the results based on the MPTbreeze treatment, which effectively gives two-loop predictions. For reference, linear theory prediction are depicted as dotted lines.
\label{fig:pkreal_MPTbreeze}}
\end{figure*}

Having confirmed an accurate power spectrum calculation with numerical PT kernels, we demonstrate our method to the resummed PT calculations. The resumed PT scheme considered here is the multi-point propagator expansion proposed by 
Ref.~\cite{Bernardeau:2008fa}. The scheme is applied to a practical power spectrum calculation at two-loop order in Refs.~\cite{Taruya:2012ut,Crocce:2012fa}. Also in modified gravity case, Ref.~\cite{Taruya:2014faa} demonstrated the one-loop calculation in $f(R)$ gravity. In this resummed PT, the multi-point propagators are the building blocks of systematic PT expansion, which possess non-perturbative properties that can be obtained in standard PT by summing up infinite series of PT expansions. A systematic construction of the $(p+1)$-point propagators, $\Gamma^{(p)}$, is thus the key in the multi-point propagator expansion, and there are methods to accurately construct propagators based on the standard PT kernels.

One proposed method is the regularized PT (RegPT) treatment in Ref.~\cite{Bernardeau:2008fa,Taruya:2012ut}. Using the standard PT kernels up to third-order, this method enables us to compute resummed power spectrum at one-loop order: 
\begin{align}
&P^{\rm RegPT}_{\delta\delta}(k)=
\left\{\Gamma^{(1)}_{{\rm reg},\delta}(k)\right\}^2\,P_0(k) 
\nonumber\\
& \quad + 2\,\int \frac{d^3\bfq}{(2\pi)^3}\,
\left\{\Gamma_{{\rm reg},\delta}^{(2)}(\bfq,\bfk-\bfq)\right\}^2\,
P_0(q)P_0(|\bfk-\bfq|),
\label{eq:pk_dd_regpt}
\end{align}
where the regularized propagators $\Gamma^{(1)}_{{\rm reg},a}$ 
and $\Gamma^{(2)}_{{\rm reg},a}$ consistent with one-loop calculation are 
respectively given by \cite{Taruya:2014faa} 
\begin{align}
&\Gamma_{{\rm reg},\delta}^{(1)}(k)=
\Biggl\{F_1(k)\left(1+ \frac{k^2\sigmad^2}{2}\right)
\nonumber\\
&\qquad \qquad+3\int\frac{d^3\bfq}{(2\pi)^3}\,F_3(\bfq,-\bfq,\bfk)P_0(q)
\Biggr\}e^{-k^2\sigmad^2},
\label{eq:Gamma1_reg_delta}\\
 &\Gamma_{{\rm reg},\delta}^{(2)}(\bfk_1,\bfk_2)=
 F_2(\bfk_1,\bfk_2)\,e^{-k^2\sigmad^2}
 \label{eq:Gamma2_reg_delta}
\end{align}
with the quantity $\sigmad$ defined by
\begin{align}
\sigmad^2=\int\frac{dq}{6\pi^2}\,\left\{G_1(q)\right\}^2\,P_0(q).
\end{align}
Since the integrals involved in the expression 
are mostly the form similar to what we saw in Eq.~(\ref{eq:pk_dd_spt}), 
the cost to numerically compute PT kernels as well as to evaluate integrals 
remains the same as in the standard PT calculations. Thus, the kernel data stored for standard PT calculation can be directly applied to the RegPT calculation. Note that in modified gravity cases, the exponential damping factor generally receive some corrections associated with nonlinear screening mechanism, but impact of this corrections is shown to be negligible in $f(R)$ gravity at the scales of our interest \cite{Taruya:2014faa}.

Fig.~\ref{fig:pkreal_MPTbreeze} presents the one-loop power spectra obtained from the RegPT treatment (long-dashed) in GR (left) and $f(R)$ gravity (right). The results are compared with standard PT results (short-dashed) and linear theory predictions (dotted). Here, we also present the power spectra data measured from $N$-body simulations, which are taken from Ref.~\cite{Taruya:2014faa}. Because of the exponential damping factor in the propagators, a large suppression of power spectrum appears at relatively low-$k$ in the one-loop results, and the agreement with simulation is mostly comparable to that of the standard PT results. Nevertheless, a crucial point is that with the damping behavior, RegPT can capture the major trend of the nonlinear smearing in the acoustic signature of power spectrum, successfully reproducing quite well the acoustic peak seen in the correlation function. By contrast, standard PT fails to compute the correlation function because of the bad high-$k$ behavior.

As another interesting example, we consider a systematic construction of propagator in Ref.~\cite{Crocce:2012fa} called MPTbreeze, with which we can develop two-loop calculations using the PT kernels up to third order. 
The power spectrum of MPTbreeze at two-loop order is given by
\begin{align}
&P_{\delta\delta}^{\rm MPT}(k)=\Bigl\{\Gamma^{(1)}_{{\rm MPT},\delta}(k)\Bigr\}^2\,P_0(k)
\nonumber\\
&\qquad +2\int \frac{d^3\bfq}{(2\pi)^3}\Bigl\{\Gamma^{(2)}_{{\rm MPT},\delta}(\bfq,\bfk-\bfq)\Bigr\}^2\,P_0(q)P_0(|\bfk-\bfq|)
\nonumber\\
&\qquad+
6\int \frac{d^3\bfp d^3\bfq}{(2\pi)^6}\Bigl\{\Gamma^{(3)}_{{\rm MPT},\delta}(\bfp,\bfq,\bfk-\bfp-\bfq)\Bigr\}^2\,
\nonumber\\
&\qquad\qquad\qquad\qquad\quad \times\,P_0(p)P_0(q)P_0(|\bfk-\bfp-\bfq|)
\label{eq:pk_MPTbreeze}
\end{align}
with the propagators:
\begin{align}
& \Gamma^{(n)}_{{\rm MPT},\delta}(\bfk_1,\cdots,\bfk_n)=F_n(\bfk_1,\cdots,\bfk_n)\,
\nonumber\\
& \quad\qquad\qquad\times\,\exp\left[\frac{3\int\frac{d^3\bfq}{(2\pi)^3} F_3(\bfq,-\bfq,\bfk)\,P_0(q)}{F_1(k)}\right].
\label{eq:prop_MPTbreeze}
\end{align}
Eq.~(\ref{eq:pk_MPTbreeze}) involves the six-dimensional integral, for which we evaluate with Monte Carlo technique \cite{Hahn:2004fe}.\footnote{The six-dimensional integral involves the symmetrized kernel $F_3$. Unlike the previous cases, it cannot be tabulated as the three-dimensional array, and this may require a bit costly numerical integration.} Note that the MPTbreeze treatment with Eqs.~(\ref{eq:pk_MPTbreeze}) and (\ref{eq:prop_MPTbreeze}) has been originally proposed and applied to the power spectrum in the $\Lambda$CDM model (i.e., GR). Nevertheless, the propagators constructed with this treatment similarly behave like what is obtained from RegPT, and we may apply MPTbreeze to the power spectrum calculations in modified gravity models.

In Fig.~\ref{fig:pkreal_MPTbreeze}, the power spectra obtained from the MPTbreeze treatment are plotted in solid lines. With the two-loop order calculations, the agreement with simulations is improved in both GR and $f(R)$ gravity, and the MPTbreeze results reproduce the power spectrum well in a range wider than RegPT and standard PT. A close look at the BAO feature reveals that the MPTbreeze tends to predict a more pronounced acoustic signal, and slightly over-predicts the simulations (see third peak at $z=2$ or second bump at $z=1$). This would be partly due to the incomplete mode-coupling treatment in constructing the multi-point propagators \cite{Taruya:2012ut}. In particular, compared to the RegPT treatment, the two-point propagators in MPTbreeze, $\Gamma^{(1)}_{\rm MPT}$, ignore the two-loop corrections, which are known to slightly reduce the amplitude of propagators \cite{Bernardeau:2012ux}. This would be a source of small discrepancy. Despite a small flaw, MPTbreeze outperforms RegPT one-loop, and with the PT kernels up to the third-order, it would give an efficient PT prediction beyond one-loop calculations.

\subsection{Transient from Zel'dovich initial condition}
\label{subsec:ZAIC}

\begin{figure}[t]
\hspace*{-0.4cm}
\includegraphics[width=9.5cm]{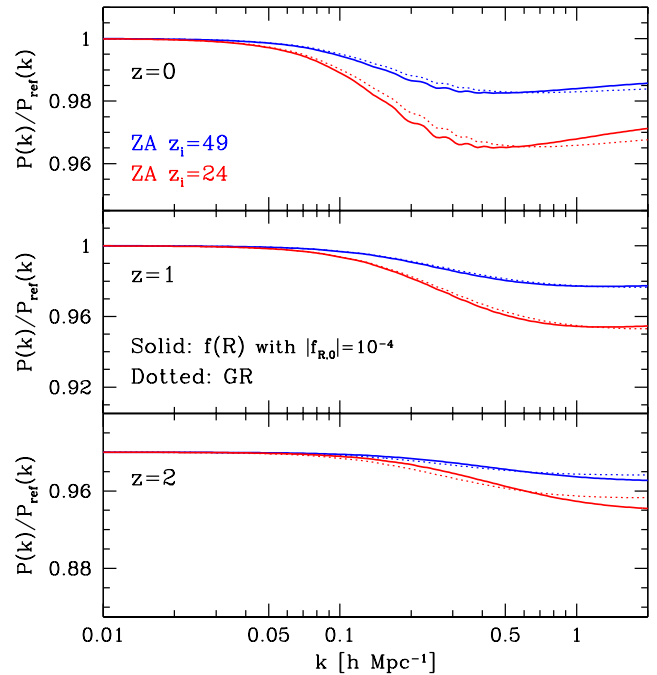}

\caption{Impact of transient from Zel'dovich initial condition on the matter power spectrum. Based on the standard PT calculation at one-loop order, impact of transient is estimated with two different initial redshifts (ZA $z_{\rm i} = 49$ and ZA $z_{\rm i} = 24$, from top to bottom in each panel). The results are normalized to the reference spectra, $P_{\rm ref}(k)$, computed with the growing-mode initial condition: $z = 0$ (top), $z = 1$ (middle) and $z = 3$ (bottom). The solid and dotted lines respectively represent the results in $f(R)$ gravity and GR. 
\label{fig:pk_ZA_transient}}
\end{figure}

So far, the numerical PT treatment has been performed with the initial conditions dominated by the linear growing-mode solution in Sec.~\ref{subsec:IC}. As we mentioned, the present numerical scheme is not only applicable to such a case, but also relevant to general initial conditions. Here, as an interesting example, we examine the power spectrum calculation with Zel'dovich initial condition. 

The Zel'dovich approximation serves as a relevant initial condition close to the one dominated by the linear growing-mode and has been frequently used in the cosmological $N$-body simulations. However, a small deviation of Zel'dovich dynamics from the growing-mode linear perturbation is known to excite a long-lived transient which can affect the statistical properties of density and velocity fields. The impact of this transient is characterized by the initial redshift $z_{\rm i}$, and has been investigated in detail both with simulation and PT in standard cosmological model (e.g., \cite{Crocce:2006ve,Scoccimarro:1997gr}), although little work has been done in the modified gravity models because of the complexity and time-consuming numerical simulation in the presence of nonlinear scalar field. Here, based on the PT calculations, we shall evaluate the impact of transient in $f(R)$ gravity, and the results are compared with those obtained in GR.

In the PT treatment, the impact of Zel'dovich transient on the late-time statistical properties is investigated by replacing the initial condition in Eq.~(\ref{eq:initial_nth}) with \cite{Crocce:2006ve,Scoccimarro:1997gr}
\begin{align}
& F_n(\bfk_1,\cdots,\bfk_n;a_{\rm i})=a_{\rm i}^n\,\widetilde{F}_n^{\rm ZA}(\bfk_1,\cdots;\bfk_n), 
\label{eq:Fn_ZA} \\
& G_n(\bfk_1,\cdots,\bfk_n;a_{\rm i})=a_{\rm i}^n\,\widetilde{G}_n^{\rm ZA}(\bfk_1,\cdots;\bfk_n). 
\label{eq:Gn_ZA}
\end{align}
where the kernels $\widetilde{F}_n^{\rm ZA}$ and $\widetilde{G}_n^{\rm ZA}$ are the symmetrized PT kernels in the Zel'dovich dynamics, and are explicitly given below up to the third order \cite{Scoccimarro:1995if,Scoccimarro:1997gr}: 
\begin{align}
&\widetilde{F}_2^{\rm ZA}(\bfk_1, \bfk_2) = \frac{1}{4}\Bigl\{
\alpha(\bfk_1,\bfk_2)+\alpha(\bfk_2,\bfk_1) + \beta(\bfk_1,\bfk_2)\Bigr\},
\nonumber\\
&\widetilde{G}_2^{\rm ZA}(\bfk_1, \bfk_2) = \frac{1}{2}\beta(\bfk_1,\bfk_2),
\nonumber\\
&\widetilde{F}_3^{\rm ZA}(\bfk_1, \bfk_2, \bfk_3) = \frac{1}{9}\Bigl\{
\alpha(\bfk_1,\bfk_{23})\,\widetilde{F}_2^{\rm ZA}(\bfk_2,\bfk_3)
\nonumber\\
&\qquad\qquad+\alpha(\bfk_{23},\bfk_1)\,\widetilde{G}_2^{\rm ZA}(\bfk_2,\bfk_3)+ \mbox{2\,\,perm}\Bigr\}
\nonumber\\
&\qquad\qquad+\frac{1}{18}\Bigl\{\beta(\bfk_1,\bfk_{23})\widetilde{G}_2^{\rm ZA}(\bfk_2,\bfk_3)+\mbox{2\,\,perm}\Bigr\},
\nonumber\\
&\widetilde{G}_3^{\rm ZA}(\bfk_1, \bfk_2, \bfk_3) = \frac{1}{6}\Bigl\{
\beta(\bfk_1,\bfk_{23})\,\widetilde{G}_2^{\rm ZA}(\bfk_2,\bfk_3)+ \mbox{2\,\,perm}\Bigr\}.
\nonumber
\end{align}

Adopting the initial conditions for PT kernels given above, we create the new kernel data, from which we compute the standard PT power spectrum in Eq.~(\ref{eq:pk_dd_spt}). The results are then divided by those obtained with the standard growing-mode initial conditions in Sec.~\ref{subsec:IC}. Fig.~\ref{fig:pk_ZA_transient} plots the output results at $z=0$, $1$ and $2$ (from top to bottom). Two different colors indicate the different initial redshifts: $z_{\rm i}=24$ (red) and $z_{\rm i}=49$ (blue). Solid lines are the results in $f(R)$ gravity, which are compared with those in GR (dotted). Note that despite the differences in cosmological parameters, the results in the GR case remarkably agree with those in Ref.~\cite{Crocce:2006ve} (see dotted lines of their Fig.~6). 

Fig.~\ref{fig:pk_ZA_transient} implies that the impact of the transients in $f(R)$ gravity is almost at the same level as seen in GR. Since the modification of gravity becomes negligible at higher redshifts and a noticeable difference appears only at $z\lesssim 2$, the results look quite reasonable. Although the standard PT is applicable to a certain narrow range in $k$, it is shown to quantitatively explain the overall trend of the transients in the GR simulations \cite{Crocce:2006ve}, and we thus expect that the results in $f(R)$ gravity is also the case. One important implication and/or remark in modified gravity is that the impact of the transients resembles that of the nonlinear screening effect on the power spectrum. Since the screening effect can affect the power spectrum even at the large scales of our interest (e.g., \cite{Oyaizu:2008tb,Koyama:2009me,Zhao:2010qy}), a precision control of the $N$-body simulation is rather crucial in modified gravity in order to discriminate the impact of screening effect from Zel'dovich transients. In this respect, the present PT calculations provide a helpful guideline to investigate this issue.

\subsection{Application to redshift-space distortions}
\label{subsec:RSD}

Since the present numerical scheme directly reconstructs the PT kernels as building block of PT, a number of applications other than presented is still possible. Here, as final remark,  we comment on the application to the redshift-space distortions (RSD). The effect of RSD is inevitable for spectroscopic measurements of galaxy clustering, and it has to be taken into account for a proper comparison to the observations. The RSD is accounted simply for mapping from real to redshift spaces through $\bfs=\bfr+v_z \hat{\bfz}/(aH)$, where the vectors $\bfr$ and $\bfs$ respectively indicate the real- and redshift-space positions, $v_z$ is the line-of-sight component of peculiar velocity, and $\hat{\bfz}$ is the unit vector parallel to the line-of-sight direction. Despite its concise expression, modeling the RSD effect on power spectrum is nontrivial due to the nonlinear nature of mapping formula, and both the non-Gaussianity and cross talk with small-scale clustering need to be incorporated into model of RSD in a proper manner. 

Among various improved RSD models recently proposed, one PT-based model of redshift-space power spectrum is given by \cite{Taruya:2010mx,Nishimichi:2011jm,Taruya:2013quf}
\begin{align}
P^{\rm(S)}(k,\mu)&=e^{-(k\mu\sigmav)^2}\Bigl\{P_{\delta\delta}(k)+2\mu^2P_{\delta\theta}(k)+\mu^4P_{\theta\theta}(k) 
\nonumber\\
&+ A(k,\mu)+B(k,\mu)\Bigr\},
\label{eq:TNS_model}
\end{align}
where $\mu$ is the directional cosine defined by $\mu=\bfk\cdot\hat{\bfz}/|\bfk|$, and $\sigmav$ is a free parameter accounting for the non-perturbative suppression due to the coherent and small-scale virialized motions. In the parenthesis, while the first three terms represent the nonlinear generalization of Kaiser term (e.g., \cite{Kaiser:1987qv,Scoccimarro:2004tg,Hamilton:1997zq}), the $A$ and $B$ terms are the next-to-leading order corrections coming from the systematic expansion of the exact power spectrum expression, and are expressed as the integrals of the bispectra and square of power spectra (see \cite{Taruya:2010mx,Nishimichi:2011jm,Taruya:2013quf} for explicit expressions). The model given in Eq.~(\ref{eq:TNS_model}) has been tested in both GR and $f(R)$ gravity \cite{Taruya:2010mx,Nishimichi:2011jm,Taruya:2013quf,Taruya:2014faa,Taruya:2013my,Song:2013ejh,Zheng:2016zxc}, and applied to the observations to simultaneously constrain geometric distances and growth of structure \cite{Oka:2013cba,Linder:2013lza,Beutler:2013yhm}. 

In similar manner to the real-space power spectrum, provided the kernels up to the third order, we can evaluate Eq.~(\ref{eq:TNS_model}) at one-loop order\footnote{Note that the $A$ and $B$ terms appear at higher order, and thus the tree-level calculations are sufficient for the bispectra and power spectra in these terms}. In Ref.~\cite{Song:2015oza}, based on Eq.~(\ref{eq:TNS_model}), the numerical PT treatment has been applied to the computation of the redshift-space correlation function in $f(R)$ gravity . Applying it to the anisotropic galaxy clustering data, a robust constraint on the model parameter, $|f_{R,0}|$, was obtained. We do not repeat the PT calculations, but we note here that Eq.~(\ref{eq:TNS_model}) does not assume any underlying theory of gravity, and can be applied to any modified gravity (see Bose \& Koyama along the line of this direction). With the present PT scheme, we uncover a variety of modified gravity models and will be able to perform a specific but a more severe test of gravity beyond the consistency test of GR.

\section{Summary}
\label{sec:summary}

In this paper, we presented a simple but powerful scheme to compute the perturbation theory (PT) kernels in general structure formation scenarios including modified gravity. The approach may be primitive, but it has versatile applicability to the statistical calculations of large-scale structure at weakly nonlinear regime. With the numerically reconstructed kernels up to the third order, we demonstrated the power spectrum calculations in both GR and $f(R)$ gravity based on the standard PT and resummed PT. With the MPTbreeze prescription, one can even perform a two-loop calculation, with which the prediction in a wider applicable range is made available. Further, with the numerical kernels starting with Zel'dovich initial conditions, the impact of transients on the matter power spectrum has been examined in $f(R)$ gravity. With a help of a model of redshift-space distortions, the present scheme can be also applied to the calculation of redshift-space power spectrum or correlation function as practical observables, and we commented on the cosmological analysis based on the numerical PT kernels.

Although the demonstrations presented here restrict the cases using PT kernels up to the third order, a reconstruction of higher-order PT kernels should be straightforward in principle, and implementing the parallel computation scheme, a much faster PT calculation will be made possible, further enlarging the applicability of the present method. This paper describes the numerical treatment focusing on the evolution equations of the PT kernels with a specific linear operator [Eqs.~(\ref{eq:linear_operator}) and (\ref{eq:evolution_PTkernel})]. But the methodology itself is quite general, and one can also apply to other type of evolution equations with different linear operator. The methodology might be useful to investigate an improved description of large-scale structure beyond the single-stream approximation.

\begin{acknowledgments}
The author would like to thank Takashi Hiramatsu for discussion and helpful comments, and Benjamin Bose and Kazuya Koyama for suggestions on the future applications. This work was supported by MEXT/JSPS KAKENHI Grant Number JP15H05899 and JP16H03977. 
\end{acknowledgments}

\bibliography{ref}
\bibliographystyle{apsrev}

\end{document}